\documentclass[12pt]{article}
\usepackage{amssymb}
\usepackage{amsmath,amsthm,amsfonts}  
\pdfoutput=1
\usepackage[applemac]{inputenc}  \usepackage{graphicx}
\usepackage{amssymb}
\usepackage{amsmath}
\usepackage{amsthm}
\usepackage{color}

\newcommand{\ket}[1]{|\kern.3ex#1\kern.3ex\rangle}
\newcommand{\bra}[1]{\langle\kern.3ex #1 \kern.3ex|}
\newcommand{\scalar}[2]{\langle\kern.3ex #1 \kern.3ex|\kern.3ex#2\kern.3ex\rangle}

\definecolor{hervecolor}{rgb}{0.8,0,0.7}

\begin{document}

\title{Position-dependent mass quantum Hamiltonians: General approach and duality
}
\author{
M.A. Rego-Monteiro$^{\mathrm{a,b}}$,   Ligia M.C.S. Rodrigues$^{\mathrm{a}}$,
E.M.F.  Curado$^{\mathrm{a,b}}$  
(\footnote{e-mail: regomont@cbpf.br,  ligia@cbpf.br,
 evaldo@cbpf.br} )\\
\emph{   $^{\mathrm{a}}$ Centro Brasileiro de Pesquisas Fisicas   } \\
\emph{   $^{\mathrm{b}}$ Instituto Nacional de Ci\^encia e Tecnologia - Sistemas Complexos}\\
\emph{  Rua Xavier Sigaud 150, 22290-180 - Rio de Janeiro, RJ, Brazil  } \\
}

\maketitle

\abstract{ {We analyze a general family of position-dependent mass quantum Hamiltonians which are not self-adjoint and include, as particular cases, some Hamiltonians obtained in phenomenological approaches to 
condensed matter physics.   
We build a general family of self-adjoint Hamiltonians  which are quantum mechanically equivalent to the non self-adjoint proposed ones. Inspired in the probability density of the problem, we construct an ansatz for the solutions of the family of self-adjoint Hamiltonians. We use this ansatz to map the solutions of the time independent Schr\"odinger equations generated by the non self-adjoint Hamiltonians into the Hilbert space of the solutions of  the respective dual self-adjoint Hamiltonians. This mapping depends on both the position-dependent mass and on a function of position satisfying a condition that assures the existence of a consistent continuity equation.  We identify the non self-adjoint Hamiltonians here studied  to a very general family of Hamiltonians proposed in a seminal article of Harrison \cite{harri} to describe varying band structures in different types of metals.  Therefore, we have self-adjoint Hamiltonians that correspond to the non self-adjoint ones found in Harrison's article. 
We  analyze three typical cases by choosing a physical position-dependent mass and a deformed harmonic oscillator potential . We completely solve the Schr\"odinger equations for the three cases; we also find and compare their respective energy levels.

\tableofcontents

\section{Introduction}
\label{intro}  
The problem of electron tunneling in systems where the band structure depends on the position, like in semiconductors, began to be treated in the early sixties \cite{harri}; later it was proposed that this variation is simulated by  a position-dependent effective mass in the one-electron Hamiltonian \cite{bendani}, and in the Hamiltonian describing graded mixed semiconductors \cite{gora}. From this time on the position-dependent mass Hamiltonians were studied in many articles in a wide range of areas other than electronic properties of semiconductors \cite{bast}, \cite{zhu}, \cite{vonroos}, \cite{young}, \cite{geller}, \cite{sinha}, like, for example, quantum wells and quantum dots  \cite{pharri}, \cite{kesha}, \cite{einevoll}, polarons \cite{zhaoliang}, etc. 

In most of those articles the choice of the position-dependent mass (PDM) Hamiltonians was guided by the characteristic of being self-adjoint, in the sense that the mean values of the physical quantities were consistently calculated in the associated Hilbert space with the usual integration measure. With this spirit, many PDM  Hamiltonians were proposed  and studied  \cite{gora}, \cite{bast}, \cite{zhu}, \cite{vonroos}, \cite{einevoll}, \cite{gorabook}, \cite{bastfurd}, \cite{LiKelin}. As a consequence some physically consistent and possibly relevant Hamiltonians have been discarded because they were not  self-adjoint \footnote{See, for example, equation (1) in \cite{gora}.}. 

In the last decades PDM Hamiltonians have also been theoretically treated in a number of articles. The interest was directed to issues like non-self-adjointness \cite{jiang2005}, solutions of the corresponding Schr\"odinger equations \cite{plastino}, \cite{jiang2005}, \cite{quesnebag}, \cite{ball}, \cite{costa}, \cite{regonobre}, \cite{monteiro2013}, \cite{dekar}, \cite{chris}, \cite{demetrios}, \cite{nikitin},  \ ordering ambiguity \cite{mustafa}, coherent states \cite{ruby} and application to some particular systems, like, for example the Coulomb problem \cite{quesnetka}.  More recently, the issue of the PDM Hamiltonians which were not within the standard self-adjoint class mentioned above were analyzed following a different approach \cite{ball}, \cite{costa}, \cite{regonobre}.   An approach to consistently quantize a non-linear system \cite{tsallis}  was recently developed. In this approach it was necessary to introduce an additional independent field which is the analog of the complex conjugate field for standard linear quantum systems. 

In this paper we study a family of linear PDM Hamiltonians  and show that the problem of self-adjointness is completely solved under certain conditions. We depart from the approach of two independent fields and define a connection between  the two fields through a mapping that depends on the position dependent mass $m(x)$ and of a function $g(x)$. In order to have appropriately well-defined probability and current densities that satisfy a continuity equation, $g(x)$ must obey a condition that depends on the form of the Hamiltonian. We show that our general non self-adjoint Hamiltonians can be identified with the very general family of Hamiltonians proposed by Harrison \cite{harri} to calculate wave functions in regions of varying band structure in superconductors, simple metals and semimetals.  Inspired in the form of the probability density,  we propose then an ansatz  that takes the solutions $\Psi(x)$ of the time independent  Schr\"odinger equations for the original non self-adjoint Hamiltonian into  new wave functions $\Omega(x)$. The wave functions $\Omega(x)$ are the solutions of the  dual Hamiltonians which are self-adjoint with the usual inner product and quantum mechanically equivalent to the original non self-adjoint ones.  We also define an inner product for the solutions $\Psi(x)$ with a generalized measure that is a function of $m(x)$ and $g(x)$.
We study three different examples of the proposed family of PDM Hamiltonians. All of them belong to the  Harrison's family of Hamiltonians \cite{harri}. For these cases we obtain the respective  dual self-adjoint Hamiltonians. In one of them the kinetic part of the dual PDM Hamiltonian belongs to the von Roos  general kinetic operator class \cite{vonroos}, but the same does not happen in the other two cases. Finally we analyze and  analytically solve the three cases taking a physical position-dependent mass and a deformed harmonic oscillator potential,  obtaining and comparing their respective energy levels.  
  
This paper is organized as follows. In section 2 we present a family of Hamiltonians with a real general potential $V(x)$ depending on a function $f(m, m^\prime)$, $m(x)$ a position dependent mass  and $m^\prime (x)$ its derivative,  and on a constant parameter $\alpha$. We obtain the  Schr\"odinger equations generated by these Hamiltonians departing from a Lagrangian density which depends on two different fields $\Psi(x,)$ and $\Phi(x)$ and on their time and spatial derivatives.  We define a transformation between these two fields that allows us to work with only one, say $\Psi(x)$, of them and to have a probability and a current density that satisfy a continuity equation. We build the quantum mechanically equivalent dual self-adjoint Hamiltonians on the Hilbert space of the solutions $\Omega(x)$ of the time independent  Schr\"odinger equations generated by them.  We define the inner products for both $\Psi(x)$ and $\Omega(x)$. 
In section 3 we analyze three different examples of the family of Hamiltonians presented, by choosing particular values for the constant parameter and for the function $f(m, m^\prime)$ and find the particular function $g(x)$. In all of them we present the particular values of the parameters that identify them with Harrison's Hamiltonians. The three examples chosen are interesting: the first one recovers a model for abrupt heterojunction between two semiconductors studied in \cite{zhu}; the other two are typical, in the sense that they introduce scales.   In these two cases the kinetic part of the dual Hamiltonian does not reduce to the von Roos general kinetic operator. In section 4 we choose a harmonic oscillator mass-dependent potential and a physically motivated particular form for $m(x)$. We solve the Schr\"odinger equations for the three cases, obtaining their corresponding eigenfunctions and energy levels. 
In section 5 we present our conclusions. 

\section{A family of general  position-dependent mass Hamiltonians}
\label{binentr}
 In this section we present a family of position-dependent-mass Hamiltonians which depend on a real function $f(m(x), m^\prime(x))$, $m(x)$ a general position dependent mass  and $m^\prime (x)$ its derivative, with a general real potential $V(x)$, given by
\begin{equation}
\label{genham}
H =  \frac{-\hbar^ 2}{2 m(x)} {\partial_x^ 2} +  \frac{\hbar^ 2}{2}  \alpha f(m(x),m^\prime (x) ) \partial_x + V(x) \,  ,
\end{equation}
where $\alpha \in \mathbb{R}$ is a dimensionless constant.  $m(x)$ is an analytical  positive function for any value of $x$.   These Hamiltonians lead to Schr\"odinger equations which are not, as will be seen, self-adjoint in the usual Hilbert space of their eigenfunctions. In what follows we show the conditions for \eqref{genham} to be self-adjoint.

\subsection{Deriving the Schr\"odinger equations from a classical Lagrangian density} 
\label{Scheq} 
When we solve the Schrodinger equation for a non self-adjoint Hamiltonian with respect to the usual inner product  we have to take into account also the adjoint equation \cite{regonobre}. Therefore our solution includes, in principle, two different fields, $\Psi(x,t)$ and $\Phi(x,t)$, and their conjugates. Note that $\Phi(x,t)$ is not the complex conjugate of $\Psi(x,t)$. 
Therefore we develop here an approach where we depart from a Lagrangian density $\mathcal{L}$ (as it was done in \cite{monteiro2013}), which depends on these 
fields, and on their time and spatial derivatives, that is
\begin{align}
\nonumber
&\mathcal{L} =  \frac{i \hbar }{2}\, \Phi(x,t)\,  \partial_t \Psi(x,t)  - 
\frac{\hbar^2}{4 m(x)} \partial_x \Phi(x,t) \partial_x \Psi(x,t) \, + \\
\nonumber
& \frac{\hbar^2}{4 }  \frac{m^{\prime}(x)}{m(x)^2} \Phi(x,t) \partial_x \Psi(x,t) - 
  \frac{\hbar^2 \alpha}{4} f(m, m^\prime)\,  \Phi(x,t) \,\partial_x \Psi(x,t) \\
  \nonumber
&-  \frac{i \hbar }{2} \Phi^ \star(x,t) \partial_t \Psi^ \star(x,t) + 
\frac{\hbar^2}{4 m(x)} \partial_x \Phi^\star(x,t) \partial_x \Psi^\star(x,t) \, + \\
& \frac{\hbar^2}{4 }  \frac{m^{\prime}(x)}{m(x)^2} \Phi^\star(x,t) \partial_x \Psi^\star(x,t) - 
  \frac{\hbar^2 \alpha}{4} f(m, m^\prime)\,  \Phi^ \star(x,t) \,\partial_x \Psi^ \star(x,t)
  \nonumber \\
  &-  \frac{1}{2} V(x) \, \Phi^ \star(x,t) \, \Psi^ \star(x,t)   - \frac{1}{2} V(x) \, \Phi(x,t) \, \Psi(x,t) \, , 
  \label{lag}
  \end{align}
  where $\star$ denotes the standard complex conjugate.
  
Using the  usual Euler-Lagrange equations for the fields $\Phi(x,t)$, $\Psi(x,t)$ and their conjugates, we straightforwardly get the following Schr\"odinger equations: 
\begin{align}
\label{1schr}
  i \hbar \partial_t \Psi(x,t)  & =   - \frac{\hbar^ 2}{2m(x)} \partial_x^ 2 \Psi(x,t)  + \frac{\hbar^ 2}{2}  \alpha f(m,m^\prime)   \partial_x \Psi(x,t) + V(x) \Psi(x,t) \\
  \label{2schr}
  -  i \hbar \partial_t \Psi^ \star (x,t)  &  =    - \frac{\hbar^ 2}{2m(x)} \partial_x^ 2 \Psi^ \star (x,t)  + \frac{\hbar^ 2}{2}  \alpha f(m,m^\prime)   \partial_x \Psi^ \star(x,t)(x,t) +
  V(x) \Psi^ \star (x,t) \, .
 \end{align}

 
\begin{align}
  \label{3schr}
  - i \hbar \partial_t \Phi(x,t)  & =   - \frac{\hbar^ 2}{2} \partial_x^ 2 \left(\frac{\Phi(x,t)}{m(x)} \right)   - \frac{\hbar^ 2 \alpha}{2} \partial_x [f(m,m^\prime) \Phi(x,t)] +  V(x) \Phi(x,t)    \\
  \label{4schr}
 i \hbar \partial_t \Phi^ \star (x,t)  & =   - \frac{\hbar^ 2}{2} \partial_x^ 2 \left(\frac{\Phi^ \star (x,t)}{m(x)} \right)   - \frac{\hbar^ 2 \alpha}{2} \partial_x [f(m,m^\prime) \Phi^ \star (x,t)]  + V(x) \Phi^ \star (x,t) \, .
 \end{align}
 Note that when the mass is a constant, \eqref{1schr} and \eqref{2schr} are the same as, respectively, \eqref{4schr} and \eqref{3schr}.

\subsection{The continuity equation} 
\label{conteq} 
We define the function $\rho(x)$ as 
\begin{equation}
\label{pd}
\rho(x,t) = \frac{1}{2 m_0} (\Psi(x, t) \Phi(x, t) + \Phi^\star(x, t) \Psi^\star(x,t)) \,  ,
\end{equation}
where $m_0$ is a mass dimensional constant. We are considering here only systems for which the integral of $\rho(x)$ over the whole space is finite. Besides, in order to be a probability density, $\rho(x)$ has to be non-negative.   This can be assured if $\Phi(x, t) = H(x) \Psi^\star(x, t)$, with $H(x) > 0$. For the sake of convenience, we choose 
$H(x) = g(x) m(x)$,  and
\begin{equation}
\label{relphipsi}
\Phi(x,t) =  g(x) m(x) \Psi^\star(x,t) \,  ,
\end{equation}
where $m(x)$ is obviously positive and we impose $g(x) > 0$. Then, \eqref{pd} becomes
\begin{equation}
\label{pd1}
\rho(x,t) = \frac{1}{m_0} (g(x) m(x) \Psi(x, t) \Psi^\star(x,t)) \,  .
\end{equation}
It is straigthforward to show that \eqref{pd1}  obeys the continuity equation
\begin{equation}
\label{continuityeq}
\partial_t \rho(x,t) + \partial_x j(x,t) = 0, 
\end{equation}
where, using Schr\"odinger equations \eqref{1schr} and \eqref{2schr}, we find the current density to be
\begin{equation}
\label{current}
 j(x,t) =  \frac{\hbar}{2 i m_0}  g(x) \left [ (\partial_x \Psi(x, t)) \Psi^\star(x,t) - \Psi(x, t) (\partial_x \Psi^\star(x,t)  \right]  \,  .
\end{equation}
This result is not valid for any function $g(x)$, but only for those obeying the condition
\begin{equation}
\label{condg}
\frac{dg(x)}{dx} = -  \alpha f(m,m^\prime)  m(x)  g(x) \,  ,
\end{equation}
which is a consequence of the continuity equation \eqref{continuityeq}. Also, is it very simple to show that \eqref{condg} makes equations \eqref{1schr} and \eqref{4schr} reduce to each other (respectively, \eqref{2schr}  and \eqref{3schr} ). This means that the class of fields $\Psi(x, t)$ and $\Phi(x, t)$ related through \eqref{relphipsi} and submitted to condition \eqref{condg} are not two independent  fields. Condition \eqref{condg} also means that given a particular Hamiltonian of the family \eqref{genham},  once we know $\alpha$ and $f(m,m^\prime)$, we have the function $g(x)$ and the probability and current densities that obey the continuity equation.

In \cite{harri}, Harrison proposed a family of Hamiltonians to describe regions of varying band structure in semiconductors, semimetals and transition metals. By comparing his current density, eq. (2) in \cite{harri},  
\begin{equation}
\label{current}
 j(x,t) =  \frac{\hbar}{2 i m_0}  \frac{\gamma}{\beta} \left [ (\partial_x \beta \phi(x, t)) \beta \phi^\star(x,t) - \beta \phi(x, t) (\partial_x \beta \phi^\star(x,t)  \right]  \,  , 
\end{equation}
with our definition of current density,  equation \eqref{current}, and identifying $\beta \phi = \psi$, we have that 
\begin{equation}\
\label{cond2}
 \frac{\gamma}{\beta} = g(x) \, . 
\end{equation}
Besides, comparing his wave equation, eq. (4) in \cite{harri}, which is the limiting case of continuous variations of the band structure, and where $\beta$,  $\gamma$ and $k_x$ are functions of position, 
\begin{equation} 
\label{harri4}
\beta \partial_x \left[ \frac{\gamma}{\beta} \partial_x [\beta \phi(x)] \right]+ \gamma  \, k_x^2 \, \beta \, \phi(x)= 0 \,   ,
\end{equation}
with the time independent Schr\"odinger equation for the Hamiltonian \eqref{genham},  $H \psi(x) = E \psi(x)$, 
with $V(x) - E = \gamma k_x^2$, 
and taking \eqref{cond2} into account, we find  
\begin{equation}
\beta = \frac{-\hbar^2}{2g(x)m(x)}
\end{equation}
and recover condition \eqref{condg}, that is, $g^\prime(x) = - \alpha f(m, m^\prime) m(x) g(x)$.

\subsection{Building the dual self-adjoint Hamiltonian}

Suggested by the form of the probability and current densities,  \eqref{pd1}  and \eqref{current}, let us define a new wave function
\begin{equation}
\label{dualta} 
\Omega(x,t) = \sqrt{g(x) m(x)} \Psi(x,t)
\end{equation} 
and, using Eq.\eqref{condg}, rewrite Eq. \eqref{1schr} for this new wave function: 
\begin{align}
\nonumber
 i \hbar \partial_t \Omega(x,t) = & -\frac{\hbar^ 2}{2 m(x)} \partial_x^2 \, \Omega(x,t) \, +  
  \frac{\hbar^ 2}{2 }  \frac{m^\prime(x)}{m(x)^2} \partial_x \, \Omega(x,t) \, - \\
 \nonumber
 & - \frac{\hbar^ 2}{4 m(x) }  \left[ - \frac{1}{2 } \alpha^2 f(m, m^\prime)^2 m(x)^2 - \alpha f(m,m^\prime) m^\prime(x) 
-\frac{3}{2}  \left(\frac{m^\prime(x)}{m(x)}\right)^2 \right.  -\\
& \left.  -\frac{1}{2} \alpha f^\prime(m,m^\prime) m(x) + \frac{m^{\prime \prime}(x)}{m(x)} \right]
 \Omega(x,t)  + V\, \Omega(x,t)
\label{hermite1}
 \end{align}

It is straightforward to show that the following Hamiltonian, defined from the above Schr\"odinger equation,

\begin{align}
\label{selfadham}
 \nonumber
&H =   -\frac{\hbar^ 2}{2 m(x)} \partial_x^2  +  
  \frac{\hbar^ 2}{2 }  \frac{m^\prime(x)}{m(x)^2} \partial_x 
  - \frac{\hbar^ 2}{4 m(x) }  \left[ - \frac{1}{2 } \alpha^2 f(m, m^\prime)^2 m(x)^2 - \right. \\ 
  & \left. \alpha f(m,m^\prime) m^\prime(x) 
-\frac{3}{2}  \left(\frac{m^\prime(x)}{m(x)}\right)^2 \right.  
 \left.  -\frac{\alpha}{2}  f^\prime(m,m^\prime) m(x) + \frac{m^{\prime \prime}(x)}{m(x)} \right]
 + V(x) 
 \end{align} 
is self-adjoint on the Hilbert space of the solutions $\Omega(x,t)$ of equation (\ref{hermite1}) with the usual inner product 
for two given solutions $\Omega_1$ and $\Omega_2$: 
\begin{equation} 
\label{inprodomega}
\langle \Omega_1(x,t), \Omega_2(x,t) \rangle \equiv \int dx \, \Omega_1^ \star (x,t) \Omega_2(x,t) \, .
\end{equation}

Relation (\ref{dualta}) can be seen as a mapping from  the solutions of Schr\"odinger equations (\ref{1schr}) and (\ref{2schr}) into  the solutions of time independent equation (\ref{hermite1}) and its Hermitian conjugate. 
Thus, motivated by (\ref{pd1}), given two solutions of equation (\ref{1schr}), namely $\Psi_1$ and $\Psi_2$, we can now define their inner  product  as
\begin{equation}
\label{inprod}
\langle \Psi_1(x,t), \Psi_2(x,t) \rangle_{gm} \equiv \int dx \, g(x) m(x) \Psi_1^ \star (x,t) \Psi_2(x,t) \, .
\end{equation}
Having the inner product above and a consistent definition of the probability density $\rho(x,t)$ \eqref{pd1} we can calculate mean values for the system described by Hamiltonian \eqref{genham}. Therefore, we have a method to  deal with the non self-adjoint Hamiltonian \eqref{genham}.
This result is valid for any analytical positive functions $m(x)$, for the functions $g(x)$ obeying  condition \eqref{condg} and for any  real potential $V(x)$. A similar result for a particular form of $m(x)$, $g(x) = 1$ and $V(x)$ was proved in theorem  1 of \cite{ball}.  

It is easy to see that the energy spectra computed from Hamiltonians (\ref{genham}) and (\ref{selfadham}) are the same; 
in the same way as it happens to the Hamiltonians, that is, the dual self-adjoint is a redefinition of the original non self-adjoint one, the physical operators will be different for the two dual systems, so that the mean values will be the same.
Thus, they are quantum mechanically equivalent. } 

This is a general result, in the sense that it is valid for  all the non self-adjoint Hamiltonians \eqref{genham} depending on a real function $f(m, m^\prime)$, with $\alpha$ a real constant and any real potential $V(x)$, provided  the two fields $\Psi(x,t)$ and $\Phi(x,t)$ are related by \eqref{phipsi} and $g(x)$ obeys condition \eqref{condg}. Besides, this is a general method to find the dual self-adjoint Hamiltonians for systems described by non self-adjoint Hamiltonians who belong to the family of Hamiltonians \eqref{genham}, under the conditions just mentioned. The non self-adjointness was the reason for discarding PDM Hamiltonians which appeared in phenomenological approaches to semiconductors (see, for instance, \cite{gora}, \cite{LiKelin}).

In \ref{conteq} we showed that our Hamiltonian \eqref{genham} is identified to the family of Hamiltonians proposed by Harrison in \cite{harri}. It is important to note that with the method here presented we can find the dual self-adjoint Hamiltonian to any non self-adjoint  contained in Harrison's family. That is, given specific forms of the functions $\beta$,  $\gamma$ and  $m(x)$,  we can find  the corresponding parameters $\alpha$ and the functions $f(m, m^\prime)$ and $g(x)$ satisfying condition \eqref{condg} and therefore the quantum mechanically equivalent self-adjoint Hamiltonian \eqref{selfadham}.

A general kinetic-energy operator for a position-dependent mass $m(x)$ system was introduced by von Roos \cite{vonroos}. In one dimension this operator is written
\begin{equation} 
\label{vonroos}
T = - \frac{\hbar^ 2}{4} \left[(m(x)^ a \partial_x m(x)^ b \partial_x m(x)^ c + m(x)^ c \partial_x m(x)^ b \partial_x m(x)^ a \right] \,   ,
\end{equation}
 and the arbitrary constants $a, b$ and $c$ obey the constraint 
\newline
 $a + b + c = -1$.  
Taking this constraint into account this operator can be written
\begin{align} 
\nonumber
T =  & - \frac{\hbar^ 2}{2 m(x)} \partial_x ^ 2 +  \frac{\hbar^ 2}{2 m(x)^ {2}}  m^ \prime(x) \partial_x  + \\ 
\label{vonroos2} 
& + \frac{\hbar^ 2}{4 m(x)} \left[ - 2 (1 + a + a^ 2 + b + ab) \left(  \frac{m^ \prime(x)}{m(x)}  \right)^ 2 
+ (1 + b) \frac{m^ {\prime \prime}(x)}{m(x)}\right] \,   .
\end{align}

The comparative analysis of von Roos kinetic operator, Eq. \eqref{vonroos2}, and the kinetic part of our Hamiltonian given by Eq. \eqref{selfadham} will be performed in the examples below.

\section{Three examples} }
\label{cases}

We have so far shown that it is possible to construct a well-defined continuity equation for the general position-dependent mass Hamiltonian (\ref{genham}). 

 In this subsection we analyze  three different classes of Lagrangian (\ref{lag}) specified by different choices of $f(m,m^\prime)$ and $\alpha$, namely:
 \begin{itemize}
\item 
 Case a:  $\alpha =  0$
\item
 Case b:  $\alpha = \frac{1}{c_1m_0},  f(m, m^\prime)=   \frac{m^\prime}{m(x)}$  ,  $[\alpha] = M^ {-1}$
\item
 Case c:  $\alpha = 2 \alpha_0 c_2, f(m, m^\prime) =  \frac{1}{m(x)}$, $\alpha_0$ a constant, $[\alpha_0] = L^ {-1}$ \, . 
\end{itemize} 
Both constants $c_1$ and $c_2$ are dimensionless. These three cases are typical in the sense that the constant $\alpha$ has no scale or scales as mass or length.

\subsection{Case a:}

In case (a) the  non self-adjoint Hamiltonian \eqref{genham} becomes
\begin{equation}
\label{ham1}
H =  \frac{-\hbar^ 2}{2 m(x)} \partial_x^ 2 +   V(x) \,  
\end{equation} 

From condition \eqref{condg}, as $\alpha = 0$, we have $g^\prime(x) = 0$ and $g$ is a constant; we take it equal to $1$.  Therefore, according to \eqref{relphipsi}, 
\begin{equation}
\label{phipsi}
\Phi(x,t) = m(x) \Psi^ \star (x,t) \, .
\end{equation} 
The functions $\Psi(x,t)$ (respectively, $\Phi(x,t)$) are the solutions of the Schr\"odinger equations \eqref{1schr} (respectively, \eqref{3schr}) for the Hamiltonian \eqref{ham1}. 

In the limit $m(x) = $ constant, we recover the usual expressions for both the probability and current densities. 

The dual self-adjoint Hamiltonian \eqref{selfadham} corresponding to Hamiltonian \eqref{ham1} is then
\begin{align}
\label{selfadhama}
 \nonumber
H_a =  & -\frac{\hbar^ 2}{2 m(x)} \partial_x^ 2  + \frac{\hbar^ 2}{2 m(x)^ 2} m^ \prime(x) \partial_x  +
 \frac{\hbar^ 2}{4 m(x)} \left[     - \frac{3}{2}  \left(  \frac{m^ \prime(x)}{m(x)}   \right)^ 2  + \frac{m^ {\prime  \prime}(x)}{ m(x)} 
  \right]  \\ + 
  & V(x)  \, .  
 \end{align} 
Comparing (\ref{vonroos2}) with the kinetic part of (\ref{selfadhama}), we see that they are the same for $a = c = - 1/2$ and $b = 0$.

From  \eqref{dualta} we have in this case 
\begin{equation}
\label{aomega} 
\Omega(x,t) = \sqrt{m(x)} \Psi(x,t) \,  ,
\end{equation} 
where $\Omega(x,t)$ is the solution of the Schr\"odinger equations \eqref{hermite1} in case (a). 

Hamiltonian \eqref{ham1} was proposed in \cite{zhu}  as a model for the abrupt heterojunction between two different semiconductors and rendered self-adjoint by an empirical approach which is a particular case of the method presented here. 

As we have showed in the general case, in \ref{conteq}, the Hamiltonians of the family \eqref{genham} are equivalent to those proposed by Harrison in \cite{harri}. In this particular case, the $\beta$ and $\gamma$ functions of Harrison's Hamiltonian \eqref{harri4} are $\beta = \gamma = \frac{- \hbar^2}{2m(x)}$ and $V(x) - E= -\frac{\ \hbar^2}{2 m(x)^2} k_x^2$.

\subsection{Case b:}

 In this case  Hamiltonian \eqref{genham} has the form
\begin{equation}
\label{bham}
H =  \frac{-\hbar^ 2}{2 m(x)} \partial_x^ 2 +  \frac{\hbar^ 2}{2 c_1 m_0 m(x)}  m^\prime (x)   \partial_x + V(x) \,  .
\end{equation} 
Let us note that in this case the $\alpha= \frac{1}{c_1m_0}$ parameter has a dimension of $M^ {-1}$. 

Integrating condition \eqref{condg}, equation \eqref{relphipsi} becomes 
\begin{equation}
\label{phipsib}
\Phi(x,t) = e^ {-\frac{m(x)}{c_1 m_0}}m(x) \Psi^ \star (x,t) \, .
\end{equation} 

The dual self-adjoint Hamiltonian \eqref{selfadham} corresponding to Hamiltonian \eqref{bham} is now
\begin{align}
\label{bselfadj}
 \nonumber
H_b =  & -\frac{\hbar^ 2}{2 m(x)} \partial_x^ 2  + \frac{\hbar^ 2}{2 m(x)^ 2} m^ \prime(x) \partial_x  -
 \frac{\hbar^ 2}{2 m(x)} \left[ \frac{3 c_1^ 2 m_0^ 2 - m(x)^ 2}{4c_1^ 2 m_0^ 2 m(x)^ 2} m^ \prime(x)^ 2 + \right. \\
 & \left. \frac{m(x) - c_1 m_0}{2 c_1 m_0 m(x)} m^ {\prime  \prime}(x) \right]   +    V(x)  \, . 
 \end{align}
From \eqref{dualta}  the new functions $\Omega(x,t)$,  
\begin{equation}
\label{bomega} 
\Omega(x,t) = e^ {-\frac{m(x)}{2 c_1 m_0}}\sqrt{m(x)} \Psi(x,t) \,  ,
\end{equation} 
are the solutions of the Schr\"odinger equations for self-adjoint Hamiltonian \eqref{bselfadj}. 

 In this case, the kinetic operator of Hamiltonian (\ref{bselfadj}) does not reduce to the Von Roos general kinetic operator (\ref{vonroos}) for any particular values of the parameters. Indeed, it is easy to see that 
 \begin{equation}
\label{baham}
H_b  = H_a +  \frac{\hbar^ 2}{2} \left[\frac{m^ \prime(x)^ 2}{4 c_1^ 2 m_0^ 2 m(x)}      -  \frac{m^ {\prime  \prime}(x)}{2 c_1 m_0 m(x)}  \right] \,  ,
\end{equation}
where $H_a$ is given by equation (\ref{selfadhama}). This shows that the von Roos kinetic operator is not the most general self-adjoint kinetic operator, as it has been assumed with frequency in the literature in the last decades. \eqref{bselfadj} is a perfectly satisfactory  PDM Hamiltonian that does not fit in the von Roos proposal. 

In this particular case, the $\beta$ and $\gamma$ functions of Harrison's Hamiltonian \eqref{harri4} are $\beta = \frac{- \hbar^2}{2m(x)} e^{\frac{m(x)}{c_1 m_0}}$, $\beta = \frac{- \hbar^2}{2m(x)}$ and $V(x) - E= -\frac{\ \hbar^2}{2 m(x)^2} k_x^2$. In  fact, this also shows that the kinectic part of Harrison's Hamiltonian proposed in \cite{harri} is more general than the von Roos' one.

\subsection{Case c:} 
Finally, in case (c) the Hamiltonian (\ref{genham}) reads
\begin{equation}
\label{cham}
 H =  \frac{-\hbar^ 2}{2 m(x)} \partial_x^ 2 +  \frac{\hbar^ 2 c_2 \alpha_0}{m(x)}  \partial_x + V(x) \,  ;
\end{equation}
the relation between functions $\Psi^\star(x,t)$ and $\Phi(x,t)$ is
\begin{equation}
\label{mapc}
\Phi(x,t) =  e^ {-2 c_2 \alpha_0 x} m(x)  \Psi^\star(x, t)\,  ,
\end{equation}
and the dual self-adjoint Hamiltonian has the form
\begin{align}
\label{seladjhamc}
\nonumber
H_c =  & -\frac{\hbar^ 2}{2 m(x)} \partial_x^ 2  + \frac{\hbar^ 2}{2 m(x)^ 2} m^ \prime(x) \partial_x  -
 \frac{\hbar^ 2}{2 m(x)} \left[ \frac{3 }{4m(x)^ 2} m^ \prime(x)^ 2 - \right. \\
 & \left. \frac{1}{2 m(x)} m^ {\prime  \prime}(x) - \frac{c_2^ 2 \alpha_0^ 2}{4}\right]   +   V(x)  \, .  
\end{align} 
 Here, the parameter $[\alpha_0] = L^ {-1}$.
 
Now the solutions of Schr\"odinger equations for Hamiltonian \eqref{seladjhamc} are
\begin{equation}
\label{dualtc} 
\Omega(x,t) = e^ {-c_2 \alpha_0 x} \sqrt{m(x)} \Psi(x,t) \,\, .
\end{equation}

As in case (b) Hamiltonian \eqref{seladjhamc} can be written as 
\begin{equation}
\label{bcham}
H_c  = H_a +  \frac{\hbar^ 2 c_2^ 2 \alpha_0^ 2}{8 m(x)}  \,  ,
\end{equation}
and its kinetic part does not reduce to the Von Roos general kinetic operator,  being another example of a satisfactory PDM Hamiltonian that  is not included in the von Roos scheme. 

In this particular case, the $\beta$ and $\gamma$ functions of Harrison's Hamiltonian \eqref{harri4} are $\beta = \frac{- \hbar^2}{2m(x)} e^{2 c_2 \alpha_0 x}$, $\gamma = \frac{- \hbar^2}{2m(x)}$ and $V(x) - E= -\frac{\ \hbar^2}{2 m(x)^2} k_x^2$. 

We remark that, after the introduction of the function $\Omega(x,t)$, instead of the Lagrangian \eqref{lag} we could write  a Lagrangian only for the field $\Omega(x,t)$, whose associate Hamiltonian is self-adjoint. This procedure was done in \cite{pvp2015}. However, with that approach we can not analyze non self-adjoint operators.

\section{Position-dependent mass Hamiltonians for a deformed harmonic oscillator}
\label{defosc}
 Position-dependent mass Schr\"odinger equations have been used to describe  semiconductor heterostructures \cite{heteros1,zhaoliang,heteros3}, as well as other kind of systems \cite{pharri, kesha, einevoll}. Among 
 all the possible $m(x)$, there is strong motivation in the literature to study the case
\begin{equation}
\label{fx}
m(x) = m_0 (1 + \gamma x^2) \, , 
\end{equation} 
which may describe the $GaAs/Al_xGa_{1-x}As$ system  \cite{galbraith, LiKelin, zhaoliang}. $m_0$ is a constant with dimension of mass. 

We choose to study the Schr\"odinger equations of  the cases (a), (b) and (c), described by Hamiltonians \eqref{ham1}, \eqref{bham} and \eqref{cham},  with the position dependent mass (\ref{fx}), and the deformed harmonic oscillator potential $V(x)$ given by
\begin{equation} 
\label{vx}
V(x) = \frac{k x^2}{2 m_0}  (1 + \gamma x^2)^{-1} \, ,
\end{equation}
where $\gamma$ is a constant with dimension $L^{-2}$ which measures the departure from the usual harmonic oscillator and $k$ is the spring constant.  With this choice the Hamiltonian \eqref{ham1} is PT-symmetric, where P means parity and T is the time reversal operator. Naturally, the standard harmonic oscillator is recovered when $\gamma$ is zero. 

In all the cases presented in  section 2 it is sufficient to solve the Schr\"odinger equations for the original non-self-adjoint Hamiltonians, namely  \eqref{ham1}, \eqref{bham}  and \eqref{cham}. From their solutions we have immediately the solutions of the Schr\"odinger equations generated by the equivalent self-adjoint Hamiltonians, which are given by \eqref{selfadhama}, \eqref{bselfadj} and \eqref{seladjhamc}. 
\subsubsection{Case a:} 

The Hamiltonian \eqref{ham1} with potential \eqref{vx} is \cite{ball}
\begin{equation} 
\label{defosca}
\mathcal{H}_a =  \frac{-\hbar^ 2 }{2 m_0 (1 + \gamma x^2)} \partial_x^ 2 +  \frac{k x^2}{2 m_0 (1 + \gamma x^2)}   \, .
\end{equation}
In the stationary case, for which $\Psi(x,t) = \exp{\frac{-i E t}{\hbar}} \Psi(x)$, the time-independent Schr\"odinger equation $\mathcal{H}_a \Psi(x) = E_a \Psi(x)$ is
\begin{equation}
\label{scheq1}
(1 + \tilde\gamma y^2)^{-1} \Psi^{\prime \prime}(y) + [ \lambda - y^2 (1 + \tilde\gamma y^2)^{-1} ]\Psi(y) = 0 \, ;
\end{equation}
above, we have defined the new variable $y = (\frac{m_0 k}{\hbar^2})^{1/4} x$ and redefined $\tilde\gamma = \gamma (\frac{\hbar^2}{m_0 k})^{1/2}$ and $\lambda = \frac{2E}{\hbar \omega}$ and, as usual, the frequency is $\omega = \sqrt{\frac{k}{m_0}}$.

\noindent
The eigensolutions $\Psi_n(y)$ are the functions: 
\begin{equation}
\label{sola1}
\Psi_{na}(y) = c_a \exp\left(-\frac{1}{2} x^2 \sqrt{1- 2 \tilde\gamma E_{na} } \right) \textmd{H}_{n} \left[ y (1-2 \tilde\gamma E_{na})^{1/4}       \right] \, ,
\end{equation}
where $c_a$ is an arbitrary constant; $\textmd{H}_n[y]$ is the Hermite polynomial and  we have the energy levels 
\begin{equation}
\label{enlevela}
E_{na} = \frac{2 n+1}{4} \left[ -\tilde\gamma (2 n+1)  + \sqrt{4 + \tilde\gamma^2 (2n+1)^2}     \right] \, .
\end{equation}
$\Phi(y,t)$ and $\Omega(y,t)$ are given respectively by \eqref{phipsi}  and \eqref{aomega}, taking into account the change of variables, with $m(y) = 1 + \tilde\gamma y^2$. Note that when $\tilde\gamma = 0$, we recover the energy levels of the harmonic oscillator.

In order to guarantee that the mass is positive-definite, we see in \eqref{fx} that $\tilde\gamma \geqslant 0$ for any value of $y$.  The asymptotic behavior of the energy levels as $n$ tend to $\infty$ is then

 \begin{equation}
\label{enasymposa}
E_{na} \sim   \frac{1}{2 \tilde\gamma}  - \frac{1}{8 \tilde\gamma^ 3 n^ 2} + O(1/n^ 3)...\, ;
\end{equation}
therefore the energy value is limited by $1/2 \tilde\gamma$, which is the maximum of the potential \eqref{vx}, and the square roots in the eigensolutions \eqref{sola1} are  well defined for all the values of the energy.

\subsubsection{Case b:} 

Taking the dimensionless constant $c_1 = 1$, the Hamiltonian \eqref{bham} with potential \eqref{vx} is
\begin{equation}
\label{defoscb}
\mathcal{H}_b =  \frac{-\hbar^ 2}{2 m_0 (1 + \gamma x^2)} \partial_x^ 2 +  \frac{\hbar^ 2}{2 m_0 (1 + \gamma x^2)}  m^\prime (x)   \partial_x + \frac{k x^2}{2 (1 + \gamma x^2)} \,  ,
\end{equation}

\noindent
With the same redefinition of variables as in case a, the eigensolutions of the time independent Schr\"odinger equations, $\mathcal{H}_b \Psi(x) = E_b \Psi(x)$,  are the functions 
\begin{equation}
\label{solb1}
\Psi_{nb}(y) = c_b \exp\left[-\frac{1}{2} y^2 \left( - \tilde\gamma + \sqrt{1+ \tilde\gamma^ 2 - 2 \tilde\gamma E_n} \right)  \right] \textmd{H}_{n} \left[ y (1 + \tilde\gamma^ 2 -2 \tilde\gamma E_n)^{1/4}       \right] \, ,
\end{equation}
and the energy levels are
\begin{equation}
\label{enlevelb}
E_{nb} = \frac{1}{4} \left[- [2 + (2 n+1)^ 2] \tilde\gamma  + (2n + 1)\sqrt{4 + \tilde\gamma^2 [8 + (2n+1)^2]}  \right]\, .
\end{equation}
$\Phi(x,t)$ and $\Omega(x,t)$ are given respectively by \eqref{phipsib}  and \eqref{bomega}, with $m(y) = 1 + \tilde\gamma y^2$. Note that when $\tilde\gamma = 0$, we recover the energy levels of the harmonic oscillator.

As $\tilde\gamma \geqslant 0$ for any value of $y$,  the asymptotic behavior of the energy levels as $n$ tend to $\infty$ is
 \begin{equation}
\label{enasymposb}
E_{nb} \sim   \frac{\tilde\gamma^ 2 + 1}{2 \tilde\gamma}  - \frac{1}{8 \tilde\gamma^ 3}(2 \tilde \gamma^ 2 + 1)^ 2  \frac{1}{n^ 2}+ O(1/n^ 3)...\, ;
\end{equation}
now the energy value is limited by $(\tilde\gamma^ 2 + 1)/2 \tilde\gamma$ which assures that  the square roots in the eigensolutions \eqref{solb1} are  well defined for all the values of the energy. 

The dual of the Hamiltonian \eqref{defoscb} is
\begin{align}
\label{selfadhamb1}
\nonumber
\mathcal{H}_b =  \mathcal{H}_a +  \frac{1}{4 m(y)} \left[ \frac{m^ \prime(y)^ 2}{2} -
 m^{\prime \prime}(y) \right]  \, ;  
 \end{align}
therefore we have an effective potential given by
\begin{equation}
\label{potef}
\mathcal{W}(y) = \frac{y^2}{2 (1 + \tilde\gamma y^2)} + \frac{1}{4 m(y)} \left[ \frac{m^ \prime(y)^ 2}{2} -  m^{\prime \prime}(y) \right] \\ .
\end{equation}
It is then easy to see that as $y \rightarrow \infty$ the limit of the potential $\mathcal{W}(y)$ is exactly the limit value of the energy,  $(\tilde\gamma^ 2 + 1)/2 \tilde\gamma$. But analyzing the potential $V(y) = y^2 (2(1 + \tilde\gamma  y^2))^{-1}$ we see that its limit is $1/2 \tilde\gamma$. Therefore in this case the threshold of the two potentials, $V(y)$ and $\mathcal{W}(y)$, are not the same. 

\subsubsection{Case c:} 
{\normalsize Following the same procedure as in Cases (a) and (b), the Hamiltonian \eqref{cham} with potential \eqref{vx}, taking $c_2 = 1$ and $\alpha_0 = 2 \sqrt{\gamma}$, is
\begin{equation}
\label{defoscc}
\mathcal{H}_c =  \frac{-\hbar^ 2}{2 m_0 (1 + \gamma x^2)} \partial_x^ 2 +  \frac{\hbar^ 2 \alpha_0}{m_0 (1 + \gamma x^2)}  \partial_x + \frac{k x^2}{2 (1 + \gamma x^2)} \,  ;
\end{equation}
 the eigensolutions are
\begin{equation}
\label{solc1}
\Psi_{nc}(y) = c_c \exp\left[2 y \sqrt{\tilde\gamma}-\frac{1}{2} y^2\sqrt{1 - 2 \tilde\gamma E_n} \right] \textmd{H}_n \left[ y (1 - 2 \tilde\gamma E_n)^{1/4}       \right] \, ,
\end{equation}
and the energy levels
\begin{equation}
\label{enlevelc}
E_{nc} = \frac{1}{4} \left[- [-8 + (2 n+1)^ 2] \tilde\gamma  + (2n + 1)\sqrt{4(1 - 4 \tilde\gamma^ 2) + \tilde\gamma^2   (2n+1)^2}  \right]\, .
\end{equation}
As $n$ tend to $\infty$ and $\tilde\gamma \geqslant 0$ for any value of $y$, \eqref{enlevelc} goes to
 \begin{equation}
\label{enasymposc}
E_{nc} \sim   \frac{1}{2 \tilde\gamma}  - \frac{1}{8 \tilde\gamma^ 3}(4 \tilde \gamma^ 2 - 1)^ 2  \frac{1}{n^ 2}+ O(1/n^ 3)...\, ;
\end{equation}
now the energy value is limited by $1/2 \tilde\gamma$ which assures that  the square roots in the eigensolutions \eqref{solc1} are  well defined for all the values of the energy. 

In this case the limit value of the effective potential of the dual Hamiltonian as $y \rightarrow \infty$  is the same  for $V(y)$, $1/2 \tilde\gamma$. 

In figure \ref{fig1} we see the behavior of the potential as a function of $x$ and the corresponding first five energy levels for the three cases presented, for $\gamma = 0.1$. 

\begin{figure}
\begin{center}
\includegraphics[width=4in]{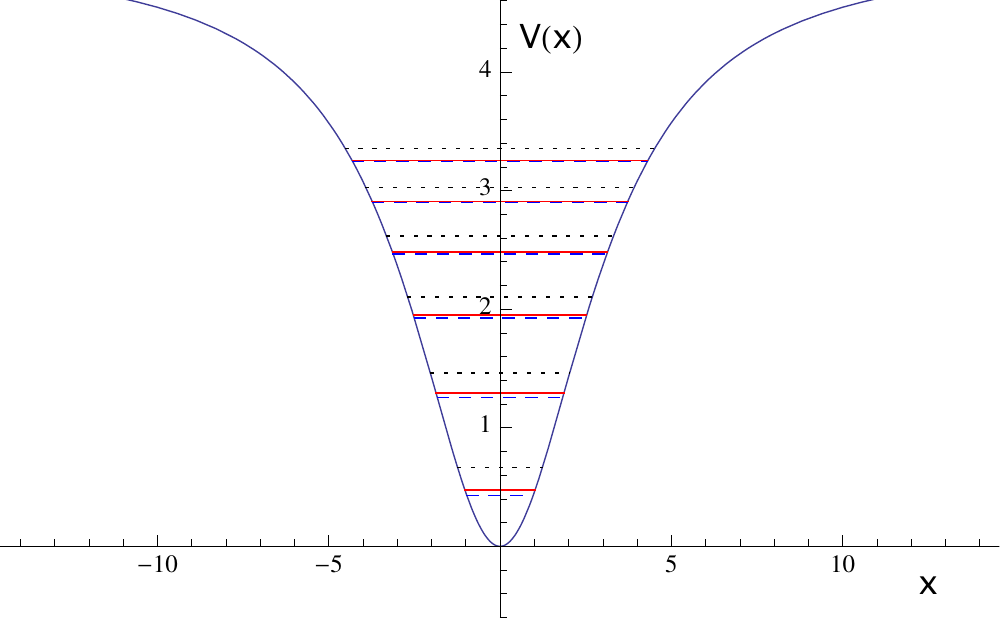} 
\caption{Behavior of the potential as a function of $x$ and the corresponding  five first energy levels for the three cases presented, for $\gamma = 0.1$. Case (a): continuous red line; case (b): dashed blue line ; case (c): dotted black line.}
\label{fig1}
\end{center}
\end{figure}

\subsubsection{Case of negative $\gamma$:}

When the deformation parameter $\gamma$ is negative, the problem can be completely solved in all our three examples of PDM Hamiltonians. A negative $\gamma$ means that in order that the mass $m(y) = 1 - |\gamma| y^ 2$ is positive definite, the potential $V(y)$ is defined only for $-1/\sqrt{|\gamma|} < y <  1/\sqrt{|\gamma|}$. Therefore there are solutions only in this region. 

As an example, in Case b the solution of the dual Hamiltonian is 
\begin{equation}
\label{solgammanegb}
\Omega_{nb}(y) = c_b \sqrt{m(y)}
\exp\left[-\frac{1}{2} y^2 \left(\sqrt{1+ \tilde\gamma^ 2 + 2 |\tilde\gamma| E_n} \right)  \right] \textmd{H}_n \left[ y (1 + \tilde\gamma^ 2 + 2 |\tilde\gamma| E_n)^{1/4}       \right] \\  ,
\end{equation}
and the energy levels are given by
\begin{equation}
\label{enasymposc}
E_{nb} = \frac{1}{4} \left[[2 + (2 n+1)^ 2] |\tilde\gamma|  + (2n + 1)\sqrt{4 + \tilde\gamma^2 [8 + (2n+1)^2]}  \right]. 
\end{equation}
It is easy to see that as $n$ tends to $\infty$ the above energy levels go to $2 |\gamma| n^2$. }

\section{Conclusions}
\label{conc}
The approach here used, which departs from a classical Lagrangian depending on two independent fields, allowed us to completely solve the question of  self-adjointness  of a class of position-dependent mass Hamiltonian systems. These Hamiltonians had been originally discarded in phenomenological approaches to semiconductors because they were  not  self-adjoint. We proved here that for a general class of these non self-adjoint Hamiltonians,  we can construct Hamiltonians which are quantum mechanically equivalent to the original ones and are self-adjoint in the usual Hilbert space of their Schr\"odinger equations solutions. This can be done if 
some function $g(x)$ that appears in consistent definitions of the probability and current densities, is restrained by the particular form of the non self-adjoint Hamiltonians.   By consistent we mean that the probability density is positive and that it obeys the usual continuity equation with an appropriate definition of the current density. 

The general non self-adjoint Hamiltonian proposed by us is identified with a large family of Hamiltonians constructed by Harrison in \cite{harri} to calculate wave functions in regions of varying band structure in superconductors, simple metals and semimetals.  That is, given specific forms of the functions involved in Harrison's Hamiltonians,  we can find  the form of  our parameters  and the function $g(x)$ satisfying condition \eqref{condg}.  Therefore we obtain the quantum mechanically equivalent self-adjoint Hamiltonians, dual to the Harrison's family.

With this method we can solve many  particular cases (that is, choosing the parameter $\alpha$ and the function $f(m(x), m^\prime(x))$) of the Hamiltonian \eqref{genham} and consequently of the Harrison's Hamiltonians. We can also expect this method to be successfully applied in Hamiltonians with non self-adjoint potentials \cite{bender}.

 We have also solved the three typical cases for a deformed harmonic oscillator potential and choosing a specific form of position-dependent mass which is potentially interesting for physical applications. Besides, the kinetic energies for these cases are particular cases of the kinectic energy in the family of Hamiltonians proposed by Harrison. We have studied these cases for positive and negative values of the deformation parameter introduced in the form of the mass. Moreover, for positive values of this parameter the systems here solved present bound states solutions with an energy threshold; for negative values the systems are confined.
 
 We believe that some Hamiltonians that were discarded because of their non self-adjointness, like, for example, in \cite{gora}, could be now treated by this method.

\section*{Acknowledgments}

We would like to acknowledge the Brazilian scientific agencies CNPq, FAPERJ and CAPES for financial support. We also thank the referees for valuable comments which helped to improve the article.

\end{document}